\documentclass[preprint,authoryear,12pt]{elsarticle}

\usepackage{hyperref}

\journal{Astronomy \& Computing}

\usepackage{multirow}
\usepackage{subfigure}
\usepackage{float}
\usepackage{caption} 
\begin{document}

\begin{frontmatter}

\title{Effect of training characteristics on object classification: an application using Boosted Decision Trees}

\author[ciemat]{I. Sevilla-Noarbe\corref{corrauthor}\fnref{phone}}
\ead{ignacio.sevilla@ciemat.es}

\author[ciemat]{P.Etayo-Sotos}

\address[ciemat]{Centro de Investigaciones Energ\'eticas, Medioambientales y Tecnol\'ogicas (CIEMAT), Av. Complutense 40, 28040 Madrid, Spain}
\cortext[corrauthor]{Corresponding author}
\fntext[phone]{Corresponding author phone: +34 91 496 25 77}

\begin{abstract}
We present an application of a particular machine-learning
method (Boosted Decision Trees, BDTs using AdaBoost) to separate stars and
galaxies in photometric images using their catalog characteristics. BDTs are a well established
 machine learning technique used for classification purposes. They have been widely used
  specially in the field of particle and astroparticle physics, and we use them here in an
  optical astronomy application. This algorithm is able to improve from simple thresholding cuts on
standard separation variables that may be affected by local
effects such as blending, badly calculated background levels
or which do not include information in other bands. The improvements are
shown using the Sloan Digital Sky Survey Data Release
9, with respect to the \textit{type} photometric classifier. We obtain an improvement in
the impurity of the galaxy sample of a factor 2-4 for this particular dataset, adjusting
for the same efficiency of the selection. Another main goal of this study is to verify the effects that different input vectors and training sets
have on the classification performance, the results being of wider use to other machine learning techniques. 
\end{abstract}

\begin{keyword}
Techniques: photometric \sep Catalogs \sep Supervised learning by classification
\end{keyword}

\end{frontmatter}

\section{Introduction}

Object classification in photometric images is an important
first step in any analysis based on catalogs from such
sources, as it constitutes a fundamental tool to build the
set to be used for model comparison or parameter
estimation. In particular, for cosmological analyses, a
significant fraction of stars contaminating the galaxy sample
can change the amplitude of the galaxy power spectrum. If
this misclassified population (represented by the impurity
fraction \textit{I}) is spatially unclustered, the amplitude
of the power spectrum is changed by a factor $(1-I)^2$ and
errors must be increased to account for it, or a correction has to be applied. A well
determined clustering amplitude is key for measuring effects such as
the galaxy bias from a specific galaxy population
\citep{coup12}, understanding large-scale cosmological
effects versus a systematic stellar contamination component (see for example
\citet{thom11} and \citet{ross11}) or distinguishing
cosmological models with primordial non-Gaussianities 
\citep{gian14}.

Star-galaxy classification has been addressed using many
different morphology based cuts since the existence of
the first photographic plate surveys (\citet{macg76}, \citet{sebo79}, \citet{heyd89}, \citet{madd90})
 and with more sophisticated techniques with the advent of
 digital imaging, machine learning methods (\citet{odew92}, \citet{weir95}, \citet{mill96}, \citet{bert96})
 and exponentially increasing computational power. Most of the implementations have
addressed the problem from the morphological point of view too. Multi-band imaging surveys, 
such as the Sloan Digital Sky Survey (SDSS) or the Canada-France-Hawaii Telescope Legacy Survey (CFHTLS), have
opened up the possibility of adding color information as
input variables (henceforth termed \textit{features}) for the classifier. 
This is explored in \citet{ball06} for SDSS Data Release 3 (DR3) and in \citet{hild12} for CFHTLenS
 and to select a pure star sample for Milky Way studies using 
 SDSS DR7 in \citet{fade12}. Recently, in \citet{male13}, the authors performed
 a study in classification using Support Vector Machines with VIPERS data as training set, highlighting
 the importance of adding infrared data to enhance the classification.
  
In this paper, we investigate the usage of AdaBoost Boosted Decision
Trees as star-galaxy classifiers, and test their performance in galaxy selection
against the standard SDSS morphological selection in SDSS Data Release 9.
We use this popular flavor of decision trees to address this
issue for the first time on optical catalog information, where we have broadened 
the scope of input features, to use color and morphological
 information simultaneously. Beyond optimizing the tree parameters, the goal is to study
  the influence of color and morphological information separately, 
  and the influence of different sizes and depth of training sets, which are required by any empirical-based
classifier.  


Decision Trees (DTs) have been explored thoroughly in the past for this
purpose, as described in \citet{such05} who were the
first to apply a DT to separate objects from the SDSS-DR2.
Later, in \citet{ball06} an axis-parallel decision tree was applied, 
using almost 500k objects from SDSS-DR3 with an
extensive exploration of parameters using as input features
the colors of the objects, for the range up to $r = 20$. In \citet{vasc11} the authors
broadened the scope of this work by comparing 13 different Decision Tree algorithms
up to $r = 21$ and using SDSS DR7 as testbed, but limiting to
morphological parameters. 


Boosted Decision Trees, introduced in \citet{freu97}, 
have been used very successfully in high energy physics \citet{roe05}
including particle classification in MiniBooNE \citep{yang05}, CMS data for
identification of the Higgs particle \citep{cms12},
 AMS \citep{agui13} and Fermi \citep{acke12}.
In optical astronomy, an application has been developed to extract photometric
redshifts from imaging surveys \citep{gerd10},
outperforming implementations based on neural networks. 
They have also been used for artifact identification in
supernovae searches  \citep{bail07}.

The paper is structured as follows: in Section \ref{sec:BDT},
BDTs and the specific implementation we have used are
detailed. In Section \ref{sec:dataset}, we describe the
dataset employed, data features chosen, training, evaluation and test
sets.  In Section \ref{sec:methodology} we detail the approach
for the optimization of the tree parameters for our specific problem, i.e., obtaining high
purity galaxy samples. We show our results for the best
parameter set in Section \ref{sec:results} and we
compare the performances for different training sets and
feature selection. Then we end with some conclusions and
possible lines of future work.


\section{Boosted Decision Trees}
\label{sec:BDT}

A Decision Tree is a structured classifier
which makes step-by-step choices based on a single \textit{feature} describing
the data. A series of sequential cuts is devised to separate the data
into one of two categories: signal and background. The value of the cuts, the feature used 
and the order in which they are applied, are
established using a training set. The process continues through these \textit{nodes}
until a final node (\textit{leaf}) is reached.

The training process starts at a root node with an arbitrary choice of 
feature and value of the cut. The separation into signal and 
background is done according to this criterion and a separation power $\theta$ is evaluated. 
In this case, we use the \textit{Gini index} to determine the performance of this 
particular choice:

\begin{equation}
G = p \cdot ( 1 - p )
\end{equation}

where \textit{p} is the purity of the selected sample (whether it be
signal or background). Using the index \textit{P} for the parent node and the 
indices \textit{s} and \textit{b} for the signal and background daughter nodes,
we determine the best choice of feature \textit{and} value of the cut which maximizes:

\begin{equation}
\theta = abs(G_P - (G_s+G_b))
\end{equation}

Every input feature is scanned, using a predetermined number of cuts for each (parameter \textit{ncuts}), to 
look for the best pair at each node. Thus the configuration of the tree continues until a minimum 
number of data points in a particular node is reached (parameter \textit{nevmin}) or if the number
 of consecutive nodes reaches a predetermined maximum (parameter \textit{maxdepth}).

Decision Trees are known to be
a powerful but unstable learning method, i.e., a small change
in the training sample can translate into a large change in the tree
and the result of the classification.  In addition, a theoretically 'perfect' classification can be achieved
if the tree is allowed to develop fully so that each leaf only contains signal or background 
data points, therefore separating fully the dataset. Of course, this is only an accurate description 
of the \textit{training} set, which most probably will not be descriptive of new data, as it has
incorporated all the noise inherent to that specific data (overfitting).

Boosting is a way of enhancing the
classification performance and increasing the stability
with respect to statistical fluctuations in the training
sample, as well as to avoid the overfitting problem. 
If a training data point is misclassified in a leaf, a weight is assigned
to that data point, according to:

\begin{equation}
w = \frac{1-\epsilon}{\epsilon}
\end{equation}

where $\epsilon$ is the misclassification rate of the tree. The weight $w$ is assigned to all such
data points and a second tree is generated anew, with the original dataset using these weights instead
(well classified values keep a weight value $w=1$). The process is
iterated tens or hundreds of times (parameter \textit{ntrees}), with all the resulting
trees combined into a 'forest' to provide significantly
enhanced classification power. This is the so-called \textit{AdaBoost} technique \citep{freu97}. With this forest of
 trees at hand, the classification of a single data point is performed based on the majority
vote of the classifications done by each tree.

We have used the Toolkit for Multivariate
Analysis framework \citep{tmva07}, provided with the ROOT analysis
package \citep{root96}, widely used in high energy physics with great success. This framework has been 
used in other astrophysical applications such as the ArborZ photometric redshift
code described in \citet{gerd10}. It is specially designed
for processing the parallel evaluation and application of different multivariate
classification techniques, among which are AdaBoost Boosted Decision Trees.

\begin{figure}[ht]
\begin{center}
\includegraphics[width=0.70\textwidth]{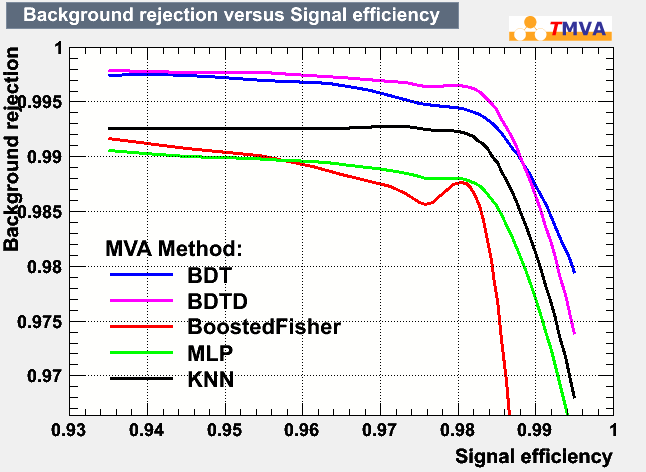}
\end{center}
\caption{Efficiency vs purity plot (ROC curve) for different machine learning methods in TMVA applied to 
a SDSS DR7 training sample described in \citet{etay12}. BDTD - decorrelated Boosted Decision Trees -
 shows the best behavior.}
\label{fig:roc_curve}
\end{figure}

A first test was performed on a training sample based on SDSS DR7 data \citep{etay12} 
 using several of the methods described in the package, with some standard, default values. 
The results are shown in Figure \ref{fig:roc_curve} via the Receiver Operator Characteristic (ROC) curve
 which measures the true positive rate versus the false positive rate of the classifier for different thresholds. 
 The BDTD method (which is a Boosted Decision Tree with a prior step of input feature decorrelation) turns 
 out to have the best performance for this problem and training set. The decorrelation step takes care of linear correlations
  between the input features (vector $\mathbf{x}$) by computing the square root $S$ of their covariance matrix and 
  constructing a new input feature vector $\mathbf{x'}=S^{-1}\mathbf{x}$. The other standard methods which were compared are:
 \begin{itemize}
 \item  k-Nearest Neighbors (\textit{kNN}): a method which searches for the \textit{k} closest training events in feature space.
 \item  Fisher Discriminant (\textit{BoostedFisher}): a linear discriminant analysis in which an axis in feature hyperspace is determined so that signal and background are as separated as possible.
 \item Neural Network  (\textit{MLP}): a multi-layer standard perceptron implementation of this classic technique, in which a non-linear mapping of the input feature vector is done onto a one-dimensional space as well. This is done through a complex mesh of cells which react to the input variables and modify their final classification accordingly.
 \end{itemize} 
This result, coupled with the success of this specific implementation in recent particle physics literature, pushed us to choose this machine learning algorithm for our tests.
  
Random Forests are a particularly successful technique too in the field of classification and regression in astronomy 
(see, e.g., \citet{carr13}). They have better generalization properties as they can account for some scatter from the training set to the application set.  On the other hand, AdaBoost BDTs can outperform slightly if the training set is representative enough. In recent tests with photometric data both give similar performances for classification (I.Sevilla-Noarbe and the DES Collaboration, in prep. and \citet{alsa15} as well as Y.Al-Sayyad, private communication). For supernovae candidate identification, random forests and boosted decision trees also compete for best performance with variable results (see \citet{bail07} or Goldstein et al. in prep.).

\section{Dataset}
\label{sec:dataset}

We have used this implementation of BDTs on Data
Release 9 (DR9) of the Sloan Digital Sky Survey (SDSS) \citep{ahn12}). The data was downloaded using the DR9
Catalog Archive Server making a selection on \texttt{modelmag\_r }
from 14 to 23 and using only spectroscopically matched
objects from the photometric table, to provide a truth value
for the purposes of evaluating the algorithm. 

Several shape features in the $r$-band and several magnitude
measurements in all bands for bands $u,g,r,i,z$ have been
used. 

We have limited the shape information to only one band
as, in first approximation, the values for these parameters across
bands should be quite compatible.  With respect to flux information, we have used a range of different
magnitude types (\texttt{fiber, model, petro, psf}) for bands $u,g,r,i,z$.

Finally, we include the photometric SDSS classification (\textit{type}) for the object, as well as the 
spectroscopic classification (\textit{class}) which we use as the reference (truth) value for 
performance in terms of purity and completeness for this work.

In Table \ref{tab:features} we summarize all the photometric catalog features used. 
The specific selection is provided in the Appendix and the resulting catalogs provides a total number of 2195172 objects.

\begin{table}
\begin{center}
\small
\caption{List of input features of the SDSS catalogs used for the training. \label{tab:features} Shape parameters taken 
only from the \textit{r} band. Color parameters include all types of magnitudes: fiber, Petrosian, model and PSF.}
\begin{tabular}{|c|c|c|}
\hline
\textbf{Parameter} &  \textbf{Type} & \textbf{Description} \\ \hline
\texttt{petroR50\_r} & Shape & \begin{tabular}[x]{@{}c@{}} Radius containing $50\%$ \\ of Petrosian flux \end{tabular}\\
\texttt{petroR90\_r} & Shape & \begin{tabular}[x]{@{}c@{}} Radius containing $90\%$ \\ of Petrosian flux \end{tabular}\\
\texttt{lnlstar\_r} & Shape & \begin{tabular}[x]{@{}c@{}} Logarithm of likelihood of fit to \\ PSF shape \end{tabular}\\
\texttt{lnlexp\_r} & Shape & \begin{tabular}[x]{@{}c@{}} Logarithm of likelihood of fit to \\ an exponential profile \end{tabular}\\
\texttt{lnldev\_r} & Shape & \begin{tabular}[x]{@{}c@{}} Logarithm of likelihood of fit to\\ a deVaucouleurs profile \end{tabular}\\
\texttt{me1\_r} & Shape & Ellipticity component 1\\
\texttt{me2\_r} & Shape & Ellipticity component 2\\
\texttt{mrrcc\_r} & Shape & Sum of second moments of object\\
\texttt{fibermag\_[ugriz]} & Magnitude & \begin{tabular}[x]{@{}c@{}} Magnitude as measured \\using the optical fiber aperture \end{tabular}\\
\texttt{petromag\_[ugriz]} & Magnitude & Petrosian magnitude in each band\\
\texttt{modelmag\_[ugriz] }& Magnitude & \begin{tabular}[x]{@{}c@{}} Magnitude as best measured by \\either exponential or deVaucouleurs profile \end{tabular}\\
\texttt{psfmag\_[ugriz]} & Magnitude & Magnitude as measured using the local PSF \\
mag\_[u]-mag\_[g] & Color & $u-g$ color \\
mag\_[g]-mag\_[r] & Color & $g-r$ color \\
mag\_[r]-mag\_[i] & Color & $r-i$ color \\
mag\_[i]-mag\_[z] & Color & $i-z$ color \\
\hline
\end{tabular}
\end{center}
\end{table}

One of the reasons for choosing this range of magnitudes and features is also to
allow for an easier comparison with the performances quoted by
\citep{vasc11}, which we have used as reference. In this case, the authors performed a
thorough testing of different Decision Tree flavors. The original AdaBoost Boosted Decision Tree implementation
 we chose is not contemplated in their study,  and we will quantitatively compare the results obtained,
 though our goal is to understand the impact of different  choices of features and training set characteristics.
 

We randomly sampled the resulting catalog into training, 
evaluation and test samples. 

\begin{itemize}
\item 200000 objects went into the
training sample to have a variety of smaller training sets
and measure training sample size dependency. From this
sample, the TMVA framework (see Section
\ref{sec:methodology}) uses a specified amount for actual
training. This is useful for the comparison of different codes in the same
execution.  
\item 800000 objects went into the evaluation sample, which we use to optimize the classifier parameters.  
\item The rest (1195172) conform the testing sample which is the one actually
used to evaluate real performance.
\end{itemize}

\begin{figure}[ht]
\begin{center}
\includegraphics[width=0.70\textwidth]{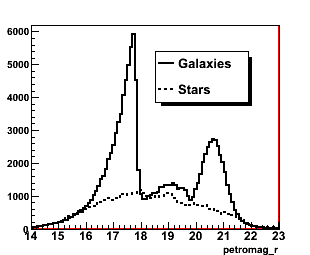}
\end{center}
\caption{Number count distribution of stars and galaxies in the training set for the downloaded SDSS DR9 catalog.}
\label{fig:dndmag}
\end{figure}

In Figure \ref{fig:dndmag} the magnitude distribution for the objects in the catalog is shown, in this case, for the 
training set (same relative distributions as for the evaluation and testing sets).

\section{Methodology}
\label{sec:methodology}

To measure performance, we define the \textit{efficiency (E)} and \textit{impurity (I)} of the
galaxy sample as:

\begin{equation}
\label{eq:eff}
E(m) = \frac{N_{galaxies}^{selected}}{N_{galaxies}^{total}}\times 100 
\end{equation}
\begin{equation}
\label{eq:imp}
I(m) =  \frac{N_{stars}^{selected}}{N_{stars+galaxies}^{selected}}\times 100
\end{equation}

where ($N_{stars}^{selected}$, $N_{galaxies}^{selected}$) is the number of true (stars, galaxies) selected 
by the classifier at magnitude $m$ and $N_{galaxies}^{total}$ corresponds to the total sample of true galaxies.
 True stars and galaxies are determined according to its spectral classification in the SDSS catalog via the \textit{class} parameter.

These metrics show a method of direct comparison against science case requirements, usually expressed in these terms (or, equivalently, completeness and purity).
In our case, we will be concerned with obtaining the lowest impurity from stars possible in our galaxy sample, given a fixed efficiency value.

To optimize and test the behavior of this classifier, we have followed these steps:
\begin{enumerate}
\item Train and evaluate on the training and evaluation sets in a grid of BDT parameters. Select best set in terms of performance (impurity level for a given efficiency).
\item Evaluate the performance on the evaluation set for different training set sizes and depths, as well as the
computation times.
\item Test the chosen configuration against the photometric \textit{type} performance provided by the 
SDSS catalogs with the test set. 
\item Verify the impact of a different choice of features assuming
the same parameters and training set size are valid. We will implicitly assume
here the independence of the BDT parameters with respect to these choices.
\end{enumerate}

The BDT parameters to be tuned are described below. The values of the grid are shown in 
Table \ref{tab:parameter_grid}, based on previous experience \citep{etay12}:

\begin{itemize}
\item \textbf{ntrees}: Number of decision trees involved in the computation.
\item \textbf{nevmin}: Minimum number of events held in a leaf.
\item \textbf{maxdepth}: Maximum size of the tree, in terms of steps from the first decision.
\item \textbf{ncuts}: Number of bins used for the cuts in each feature being tested.
\end{itemize}

\begin{table}
\begin{center}
\caption{List of Boosted Decision Trees parameters as named in the TMVA environment and the values for which their
 performance is evaluated. In bold face, the selected values after parallel coordinate analysis (see text). \label{tab:parameter_grid}}
\begin{tabular}{|c|c|}
\hline
\textbf{Parameter} & \textbf{Range} \\ 
\hline
ntrees & 200,400,1000,\textbf{2000},3000 \\
nevmin & 10,\textbf{50},100,400,1000 \\
maxdepth & 5,\textbf{10},15,20,30 \\
ncuts & 20,50,\textbf{200},500,1000 \\ 
\hline
\end{tabular}
\end{center}
\end{table}

This grid was explored by submitting multiple batch jobs to a 
cluster both for the training and evaluation sets, as defined
in Section \ref{sec:dataset}.  We have narrowed down to this particular set of values in each
case after a test in a wider range, the limits being imposed
by performance and relative gain with respect to execution time.

The computation was performed using the Euler cluster at the CIEMAT Spanish national lab, in Madrid.
This cluster is composed of 144 nodes with 2GB RAM and two Quad-core Xeon processors each
 at 3.0 GHz clock speed.
 
\section{Results and discussion}
\label{sec:results}

We will now use the impurity metric defined in equation \ref{eq:imp} as the value to compare the performance of each Boosted Decision Tree set we produce, as well as for the SDSS \textit{type} parameter. In order to compare fairly with the latter we have adjusted the selection cut for each BDT so that its efficiency (equation \ref{eq:eff}) was within $0.1\%$ of the efficiency found for the \textit{type} classifier at that particular magnitude bin (in \texttt{modelmag\_r}).

In this section we detail the strategy to select the parameters for the BDT, as well as the impact of a varying  training set size, composition, depth and feature selection, which can also provide hints for the expected performance of other machine learning methods which extract the same information and relationships in the data.

\subsection{Selection of optimal parameters for the BDTs}
\label{sec:parameter_selection}

\begin{figure}[ht]
\begin{center}
\includegraphics[width=0.90\textwidth]{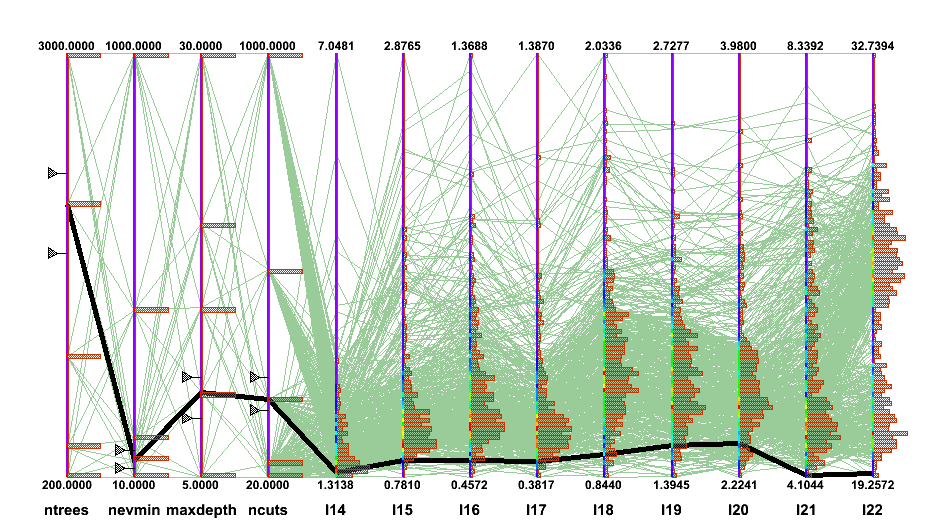}
\end{center}
\caption{Parallel coordinates visualization showing the relationships of the four input parameters (first four axes) with the impurities obtained at each magnitude range  (last nine axes). The thicker line corresponds to the choice of parameters and the resulting impurity levels on the evaluation set, chosen for this paper.}
\label{fig:parallelcoord}
\end{figure}

We have executed the 625 jobs of the parameter grid, corresponding to all combinations of parameters in  Table \ref{tab:parameter_grid}), and obtained a 9-element vector with the impurity level for each magnitude bin (\texttt{modelmag} in $r$-band). We select the parameter set which provides the best (lowest) overall
impurity level in the evaluation set (boldface in Table \ref{tab:parameter_grid}). In order to do so, we visualize the performance of all possible combinations through a \textit{parallel coordinates} plot such as the one shown in Figure \ref{fig:parallelcoord}. This type of representation allows showing a set of points in a \textit{N}-dimensional space, with four input parameters and a 9-component vector output, so that \textit{N} lines are drawn, each encompassing the whole range of each input and output value. A point in this \textit{N}-dimensional space is represented as a polyline with vertices on the parallel axes. In our case, the first four points connected by the polyline represent the input parameter values of Table \ref{tab:parameter_grid}, and  are connected by continuing the polyline to the 9-output components of the impurity vector, which is the metric we are using. With this tool, it is possible to distinguish the overall performance of a particular combination of input values for the parameters against the background of other possible combinations. In the figure, the specific combination which provides a good overall impurity level throughout the magnitude range is shown with a thicker line. This combination is highlighted in boldface in Table \ref{tab:parameter_grid}.



\begin{figure}[ht]
\begin{center}
	\subfigure{
		\includegraphics[width=0.40\textwidth]{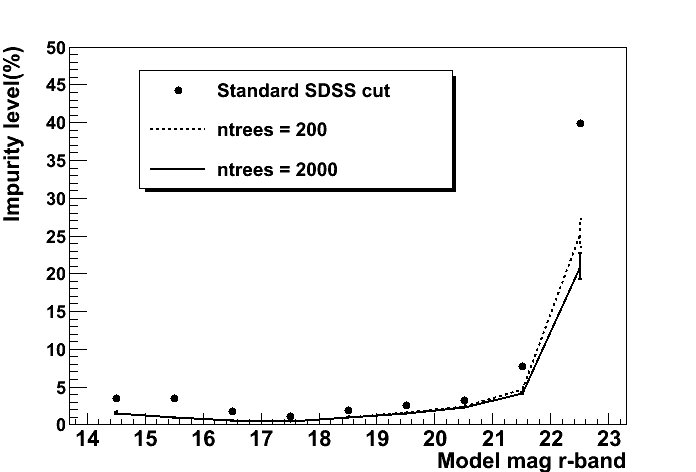}
	}
	\subfigure{
		\includegraphics[width=0.40\textwidth]{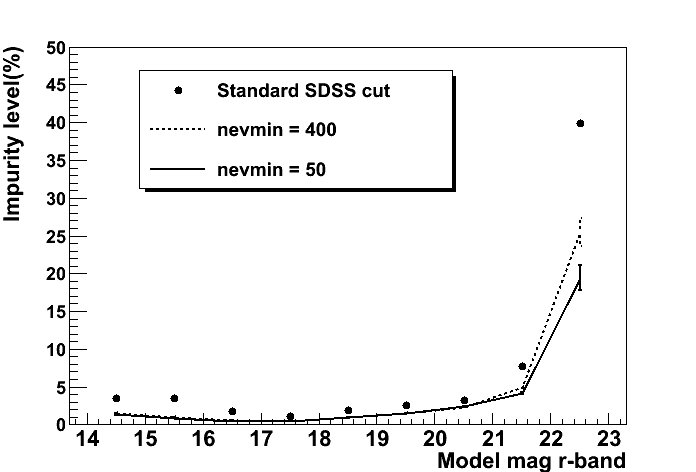}
	}
	\subfigure{
		\includegraphics[width=0.40\textwidth]{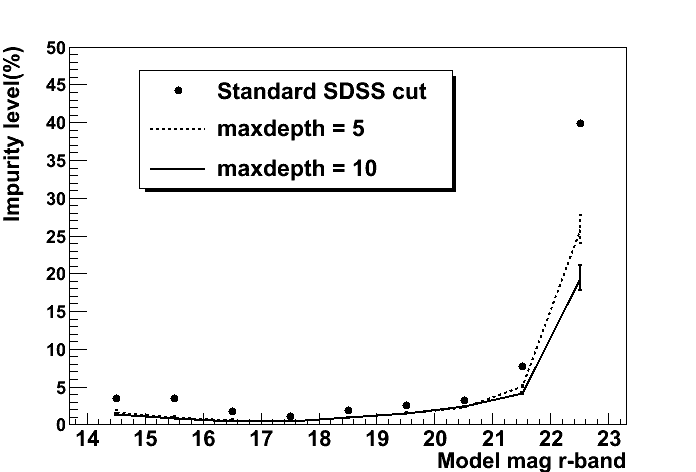}
	}
	\subfigure{
		\includegraphics[width=0.40\textwidth]{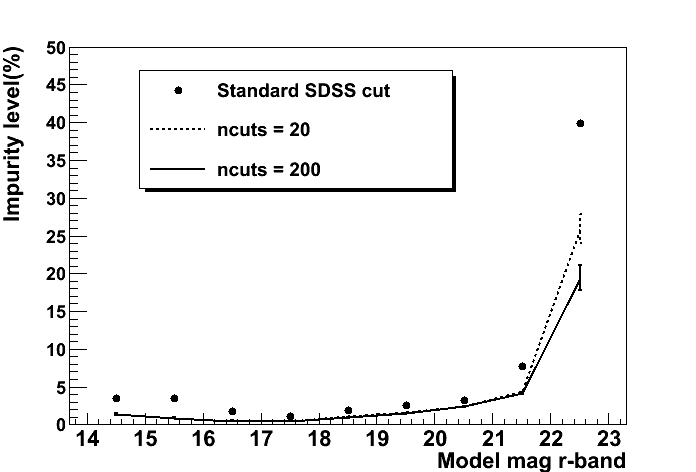}
	}
\end{center}
\caption{Impurity level comparison with variations of a single parameter.} 
\label{fig:parameter_dependence}
\end{figure}

Examples of the effect of the change of specific features are shown in Figure \ref{fig:parameter_dependence}. The increase in the number of trees, tree depth and number of cuts in each feature decrease the impurity level achieved until a certain stable value beyond which there is no significant gain though we incur in an execution time penalization as well as increasing the risk of overfitting (although the boosting approach tries to avoid this). 

The training set size used was subselected to include 30000 randomly-picked galaxies
and 6000 stars likewise chosen. This size provides a suitable trade-off between computation time and 
training performance. We will show in Section \ref{sec:size} the results when these values are modified.

\subsection{Dependence with training set size} 
\label{sec:size}

The number of samples, both for signal and background, is directly related to the performance of the classifier, as an increasingly varied array of galaxy and star types are covered. The fact that we have used an unbiased sample, in the sense that no targeting has been specifically done for these objects, and covering a wide area, which diminishes the impact of sample variance, makes this catalog an exceptional resource, up to the available depth. These characteristics allow us to understand the impact of the training sample solely in terms of its size, without worrying about the specific kinds of objects which populate the sample.

\begin{figure}[H]
\begin{center}
	\subfigure{
		\includegraphics[width=0.40\textwidth]{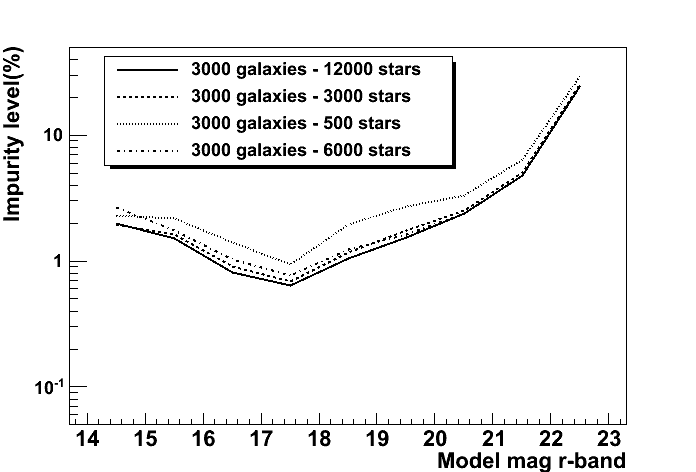}
	}
	\subfigure{
		\includegraphics[width=0.40\textwidth]{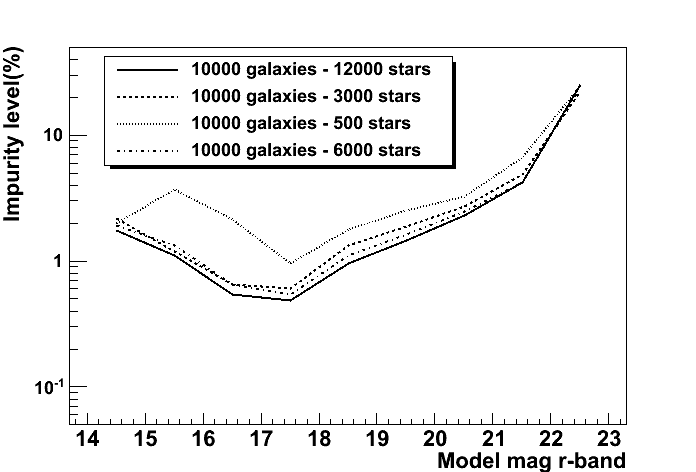}
	}
	\subfigure{
		\includegraphics[width=0.40\textwidth]{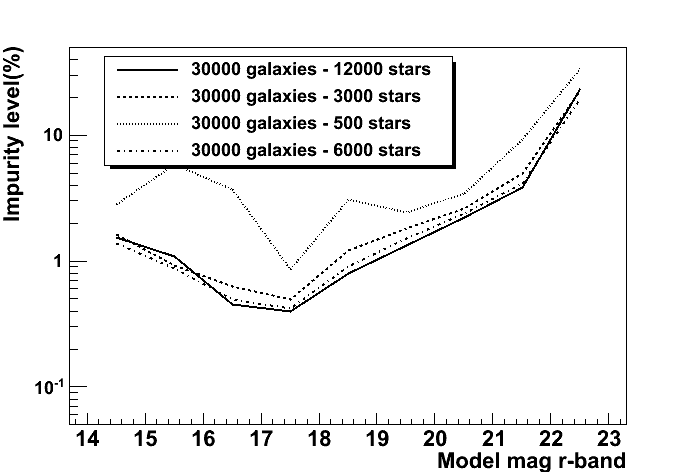}
	}
	\subfigure{
		\includegraphics[width=0.40\textwidth]{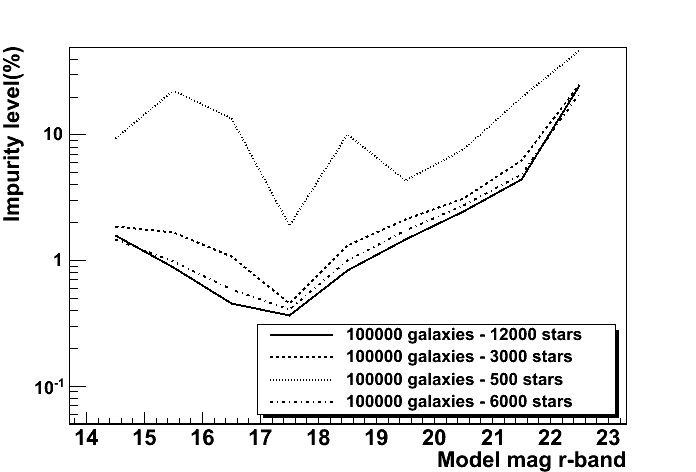}
	}
\end{center}
\caption{Impurity level for 3000 (upper left), 10000 (upper right), 30000 (lower left) and 100000 (lower right) galaxies in the training set, and variable star sample.} 
\label{fig:trainsize}
\end{figure}

In Figure \ref{fig:trainsize} we show the evolution of the impurity level (our chosen comparison metric) with respect to the size of the training set, as well as the relative mixing of galaxies and stars. In Table \ref{tab:training_times} the training times for different source and background sample sizes are shown. 

By studying both results, a good compromise in terms of speed and impurity metric is the choice of using 30000 galaxies and 6000 stars. Increasing the number of galaxies does not improve things much and on the other hand, not providing sufficient number of stars to have a well balanced sample can ruin our impurity performance, as the classifier will tend to classify objects as galaxies. This can be seen in the lower right panel of Figure \ref{fig:trainsize} or in any panel, when the available star sample is only populated by 500 objects, leaving a small star-to-galaxy ratio which will tend to make objects to be classified predominantly as galaxies, as some specific stellar types may have been randomly left out or are underrepresented.

Note well that the process of selecting the most adequate training set size and mix, as well as the one with best performance (Section \ref{sec:parameter_selection}) has been through an iterative process in which a default set of parameters were used with varying sample size, then the most adequate parameters were chosen, again training sample size sensitivity was reanalyzed, etc. 

\begin{table}[H]
\begin{center}
\caption{Training times (in seconds) for different choices of galaxies and stars, source and background respectively, in the training sample.  \label{tab:training_times}}
\begin{tabular}{|c|c|c|c|c|c|}
\hline
Training time (s) & \multicolumn{5}{|c|}{\textbf{Nb. Galaxies in training}} \\
\hline
\textbf{Nb. stars in training} & \textbf{1000} & \textbf{3000}  & \textbf{10000}  & \textbf{30000}  & \textbf{100000}   \\ \hline
\textbf{500}  & 36 & 89 & 201 & 599 & 2550  \\ \hline
\textbf{3000}  &118 & 203 & 280 & 709 & 2980 \\ \hline
\textbf{6000}  & 253 & 346 & 355 & 795 & 3030\\ \hline
\textbf{12000} & 547 & 322 & 489 & 1160 & 3230  \\ \hline
\end{tabular}
\end{center}
\end{table}

\subsection{Dependence with training set depth}

A common circumstance that many present and future photometric surveys will face is the lack of adequate training for their machine learning based classifiers, due to the unavailability of overlapping areas with spectroscopic information reaching the full survey depth. In this section, we experiment with variations in the availability of training samples depending on the magnitude limit we impose to them, and verify the impact on the impurity of the galaxy sample.

\begin{figure}[H]
\begin{center}
\includegraphics[width=0.70\textwidth]{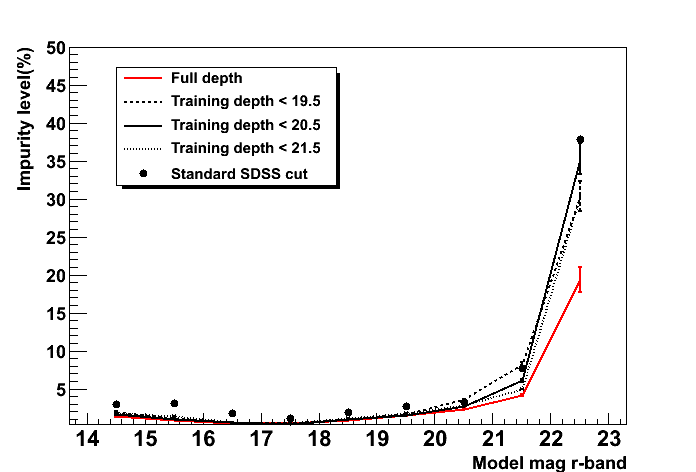}
\end{center}
\caption{Effect of the usage of different depths for the training sample, defined in terms of a limit to \texttt{modelmag\_r}.}
\label{fig:perf_vs_depth}
\end{figure}

In Figure \ref{fig:perf_vs_depth} the results for different choice of training depths are shown.

It can be verified that such variations are indeed significant and more important specially at magnitudes deeper than where training was available. In fact, a morphological cut approach, such as the SDSS photometric \texttt{type} can be as valid as a machine learning method employing multiple features, if the training is not deep enough.

\subsection{Dependence with features}

It is interesting to explore what the different features from Table \ref{tab:features} contribute to the overall result. We separate for this study the available features into shape-related, magnitude-related and color-related, as specified on the second column of the aforementioned table.

\begin{figure}[H]
\begin{center}
\includegraphics[width=0.70\textwidth]{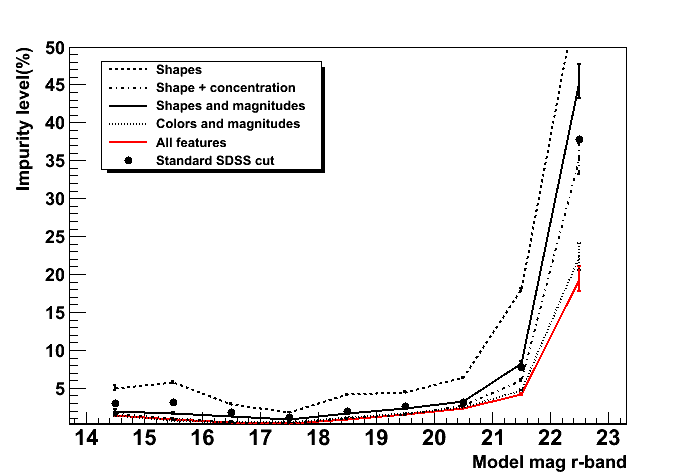}
\end{center}
\caption{Effect of the usage of different kinds of input features on the galaxy purity classification, as compared with the SDSS type parameter, based on 'flux concentration' (\texttt{modelmag - psfmag}).}
\label{fig:perf_vs_case}
\end{figure}

In Figure \ref{fig:perf_vs_case} the impact of different choice of features are shown.

Color and magnitude information are clearly the most important sources from which the BDTs cull their information. Using shape features (light radii, ellipticity, fit likelihoods to models), on the other hand, cannot compete with the clues provided by a concentration parameter, which proves to be a robust measurement. Indeed, adding this combination to the shape input features is an important improvement to the classifier, and provides a similar response.
Therefore, when using single band information, a simple cut on a 'concentration'-like parameter (e.g. differences in fluxes from PSF magnitudes and model magnitudes, or the \texttt{SPREAD\_MODEL} parameter showcased in \citet{desa12} and \citet{soum13}) should be enough.

\subsection{Evaluation against SDSS DR9 photometric type}

\begin{figure}[ht]
\begin{center}
\includegraphics[width=0.70\textwidth]{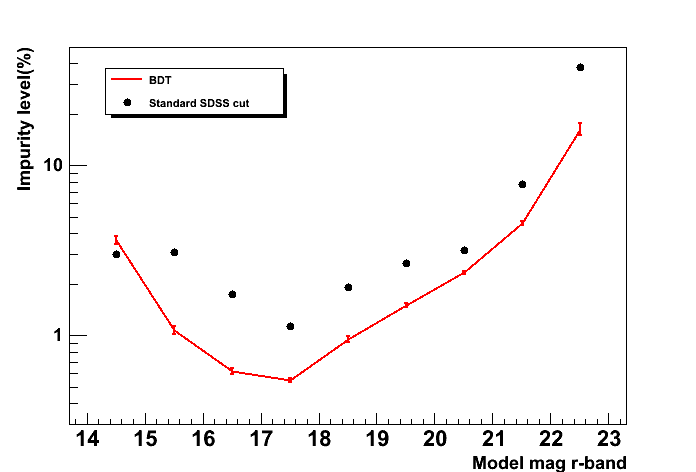}
\end{center}
\caption{Comparison of the impurity in the galaxy sample in SDSS DR9 for the standard \textit{type} classifier, and the method presented in this paper using Boosted Decision Trees.}
\label{fig:perf_vs_type}
\end{figure}

To test independently of the training and evaluation set, we have used the test set, with the chosen parameters and training size. The results are shown in in Figure \ref{fig:perf_vs_type}, to be compared with the results for the photometric \textit{type} provided with the catalog. This classifier has a similar performance to the standard \texttt{psfmag-cmodelmag} $> 0.145$ cut\footnote{https://www.sdss3.org/dr8/algorithms/classify.php}. Our suggested BDT approach provides an improvement of 2 to 4 times on the impurity level, with this relatively simple training approach, using color, magnitude and shape information.

We can compare our results with the ones reported by \citet{vasc11} by fixing the efficiency values to approximately the same ones they show in Figure 6 of their paper. We obtain impurity levels around or below $1\%$ (except of $\sim2\%$ at the magnitude bin of 14-15), maybe slightly smaller than what it is shown for their DT classifier. However, their dataset is shallower, and their choice of parameters is morphological, further evidencing the conclusions of our work in terms of dependency of performance with depth and feature selection.

\begin{figure}[ht]
\begin{center}
	\subfigure{
		\includegraphics[width=0.45\textwidth]{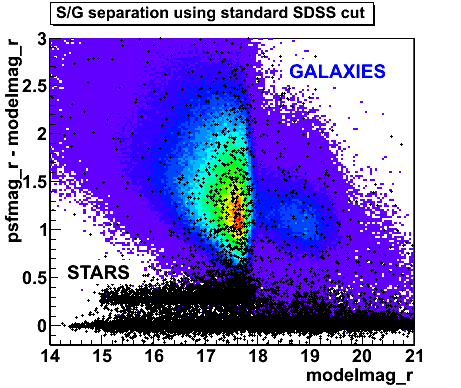}
	}
	\subfigure{
		\includegraphics[width=0.45\textwidth]{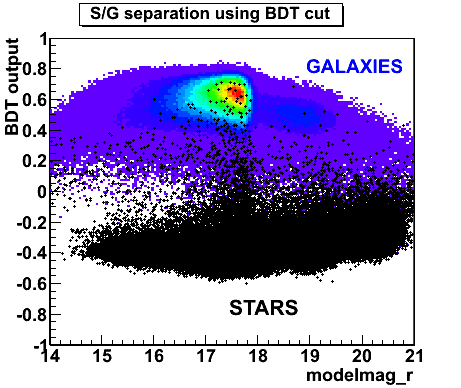}
	}
\end{center}
\caption{PSF - model magnitude (left) and BDT (right) separation variables of spectroscopic stars and galaxies, as a function of the modeled magnitude in the r band (for SDSS DR7 objects, from \citet{etay12})}
\label{fig:separators_visual}
\end{figure}

The separation power of the BDT method can be qualitatively appreciated in Figure \ref{fig:separators_visual}.
 
\section{Conclusions}

In this work we have showed the improvements that we obtain by applying AdaBoost Boosted Decision Trees on SDSS DR9 photometric data to classify objects as stars or galaxies, using colors as well as morphological features, using a prior feature decorrelation step.  This technique, very successful in other fields akin to astrophysics, has never been applied, to our knowledge, for this optical astronomy application. Using spectroscopic data from the survey itself, we have tuned the parameters for best performance, and then compared against the usage of the standard \textit{type} photometric parameter from the SDSS catalogs, obtaining up to a factor 4 improvement in the impurity of the galaxy sample.

In addition to this we have made a few variations on the training sample to verify the impact on the classification. These results have general validity for other machine learning classifiers, which rely on the same available catalog information.

The BDT parameter choice has been done scanning through different values, and using different training set sizes. Using a parallel coordinates plot for this kind of analysis has proven to be a very simple and useful tool which is widely extensible to other machine learning approaches. For this unbiased sample, which would cover a varied array of galactic and stellar types, we notice that beyond 30k objects for training, the improvement is not relevant.

Training set mixture is a desirable feature of the training set, as too many galaxies in the sample tend to fool the classifier oftentimes into assigning stars as galaxies. Therefore, a relevant presence of background objects (in our case, at least 20\%) is necessary to ensure an applicable training.

Colors and magnitudes are the most important features used by the BDTs to improve the performance over the morphology-based SDSS \textit{type}, though the latter, proves to be a simple and robust figure (based on concentration of light) which is easy to implement and can be sufficient for many studies, as has been proven in the literature. Using light concentration together with shape information in this machine learning implementation simply converges to the standard \textit{type} classification. It is the addition of information of color space that gives the additional edge.


Extensions to this work include new object types such as quasars (more relevant on next generation deeper surveys) or image artifacts. Including photometric redshift as an input feature is also an alternative avenue to pursue as a 'color' selection. Finally, a veritable improvement of this classifier would be incorporating it into a Bayesian framework. This way, the computation of correlation functions for example, that made use of the survey galaxies would not have to have a sample previously selected, but incorporate all objects with an associated probability. See for example \citet{fade12}, \citet{carr14}, Kim et al. in prep. This would be an asset too for weak lensing measurements, in which a contamination of the shear catalog by stars introduces an additive bias in the shear-shear correlation function (E. Sheldon, private communication).

The code used is made available\footnote{\texttt{http://github.com/nsevilla/BDT\_sg\_classification}} with this publication. I requires previous installation of the ROOT framework\footnote{\texttt{http://root.cern.ch}}. We used version 5.18 for all our tests. The dataset can be downloaded using the query in the Appendix and is also available online\footnote{\texttt{http://wwwae.ciemat.es/$\sim$sevilla/bdts}}.

\section{Acknowledgements}

ISN would like to thank Robert Brunner, Alex Drlica-Wagner and O.Pe\~na for useful discussions in the development and testing of this code, as well as for insights into its usage.

We thank the Spanish Ministry of Economy and Competitiveness (MINECO) for funding support through grant FPA2013-47986-C3-2-P.

\newpage

\section*{Appendix: Query to obtain the catalog}

\begin{tt}

SELECT 

p.ra, p.dec

p.petromag\_u, p.petromag\_g, p.petromag\_r, p.petromag\_i, 

p.petromag\_z, p.modelmag\_u, p.modelmag\_g, p.modelmag\_r, 

p.modelmag\_i, p.modelmag\_z, p.psfmag\_u, p.psfmag\_g, 

p.psfmag\_r, p.psfmag\_i, p.psfmag\_z, p.fibermag\_u, 

p.fibermag\_g, p.fibermag\_r, p.fibermag\_i, p.fibermag\_z,

p.petrorad\_r, p.petror50\_r, p.petror90\_r, p.lnlstar\_r, p.lnlexp\_r, 

p.lnldev\_r, p.me1\_r, p.me2\_r, p.mrrcc\_r, s.class

into mydb.SGtable

from dr9.PhotoPrimary

AS p

JOIN dr9.SpecObj

AS s

ON s.bestobjid = p.objid

WHERE p.modelmag\_r between 14.0 and 23.0

\end{tt}

\bibliographystyle{model2-names}
\bibliography{sgsep_bdt_elsevier}

\begin{thebibliography}{35}
\expandafter\ifx\csname natexlab\endcsname\relax\def\natexlab#1{#1}\fi
\providecommand{\url}[1]{\texttt{#1}}
\providecommand{\href}[2]{#2}
\providecommand{\path}[1]{#1}
\providecommand{\DOIprefix}{doi:}
\providecommand{\ArXivprefix}{arXiv:}
\providecommand{\URLprefix}{URL: }
\providecommand{\Pubmedprefix}{pmid:}
\providecommand{\doi}[1]{\href{http://dx.doi.org/#1}{\path{#1}}}
\providecommand{\Pubmed}[1]{\href{pmid:#1}{\path{#1}}}
\providecommand{\bibinfo}[2]{#2}
\ifx\xfnm\relax \def\xfnm[#1]{\unskip,\space#1}\fi
\bibitem[{{Aguilar} et~al.(2013){Aguilar}, {Alberti}, {Alpat}, {Alvino},
  {Ambrosi}, {Andeen}, {Anderhub}, {Arruda}, {Azzarello}, {Bachlechner} and
  et~al.}]{agui13}
\bibinfo{author}{{Aguilar}, M.}, \bibinfo{author}{{Alberti}, G.},
  \bibinfo{author}{{Alpat}, B.}, \bibinfo{author}{{Alvino}, A.},
  \bibinfo{author}{{Ambrosi}, G.}, \bibinfo{author}{{Andeen}, K.},
  \bibinfo{author}{{Anderhub}, H.}, \bibinfo{author}{{Arruda}, L.},
  \bibinfo{author}{{Azzarello}, P.}, \bibinfo{author}{{Bachlechner}, A.},
  \bibinfo{author}{et~al.}, \bibinfo{year}{2013}.
\newblock \bibinfo{title}{{First Result from the Alpha Magnetic Spectrometer on
  the International Space Station: Precision Measurement of the Positron
  Fraction in Primary Cosmic Rays of 0.5-350 GeV}}.
\newblock \bibinfo{journal}{Physical Review Letters} \bibinfo{volume}{110},
  \bibinfo{pages}{141102}.
\newblock \DOIprefix\doi{10.1103/PhysRevLett.110.141102}.
\bibitem[{{Ahn} et~al.(2012){Ahn}, {Alexandroff}, {Allende Prieto}, {Anderson},
  {Anderton}, {Andrews}, {Aubourg}, {Bailey}, {Balbinot}, {Barnes} and
  et~al.}]{ahn12}
\bibinfo{author}{{Ahn}, C.P.}, \bibinfo{author}{{Alexandroff}, R.},
  \bibinfo{author}{{Allende Prieto}, C.}, \bibinfo{author}{{Anderson}, S.F.},
  \bibinfo{author}{{Anderton}, T.}, \bibinfo{author}{{Andrews}, B.H.},
  \bibinfo{author}{{Aubourg}, {\'E}.}, \bibinfo{author}{{Bailey}, S.},
  \bibinfo{author}{{Balbinot}, E.}, \bibinfo{author}{{Barnes}, R.},
  \bibinfo{author}{et~al.}, \bibinfo{year}{2012}.
\newblock \bibinfo{title}{{The Ninth Data Release of the Sloan Digital Sky
  Survey: First Spectroscopic Data from the SDSS-III Baryon Oscillation
  Spectroscopic Survey}}.
\newblock \bibinfo{journal}{The Astrophysical Journal Supplements Series}
  \bibinfo{volume}{203}, \bibinfo{pages}{21}.
\newblock \DOIprefix\doi{10.1088/0067-0049/203/2/21},
  \href{http://arxiv.org/abs/1207.7137}{\tt arXiv:1207.7137}.
\bibitem[{{AlSayyad} et~al.(2015){AlSayyad}, {McGreer}, {Fan}, {Connolly},
  {Ivezic} and {Becker}}]{alsa15}
\bibinfo{author}{{AlSayyad}, Y.}, \bibinfo{author}{{McGreer}, I.D.},
  \bibinfo{author}{{Fan}, X.}, \bibinfo{author}{{Connolly}, A.J.},
  \bibinfo{author}{{Ivezic}, Z.}, \bibinfo{author}{{Becker}, A.C.},
  \bibinfo{year}{2015}.
\newblock \bibinfo{title}{{Optical Variability and Classification of High
  Redshift (3.5 < z < 5.5) Quasars on SDSS Stripe 82}}, in:
  \bibinfo{booktitle}{American Astronomical Society Meeting Abstracts}, p.
  \bibinfo{pages}{144.46}.
\bibitem[{{Bailey} et~al.(2007){Bailey}, {Aragon}, {Romano}, {Thomas}, {Weaver}
  and {Wong}}]{bail07}
\bibinfo{author}{{Bailey}, S.}, \bibinfo{author}{{Aragon}, C.},
  \bibinfo{author}{{Romano}, R.}, \bibinfo{author}{{Thomas}, R.C.},
  \bibinfo{author}{{Weaver}, B.A.}, \bibinfo{author}{{Wong}, D.},
  \bibinfo{year}{2007}.
\newblock \bibinfo{title}{{How to Find More Supernovae with Less Work: Object
  Classification Techniques for Difference Imaging}}.
\newblock \bibinfo{journal}{The Astrophysical Journal} \bibinfo{volume}{665},
  \bibinfo{pages}{1246--1253}.
\newblock \DOIprefix\doi{10.1086/519832},
  \href{http://arxiv.org/abs/0705.0493}{\tt arXiv:0705.0493}.
\bibitem[{{Ball} et~al.(2006){Ball}, {Brunner}, {Myers} and {Tcheng}}]{ball06}
\bibinfo{author}{{Ball}, N.M.}, \bibinfo{author}{{Brunner}, R.J.},
  \bibinfo{author}{{Myers}, A.D.}, \bibinfo{author}{{Tcheng}, D.},
  \bibinfo{year}{2006}.
\newblock \bibinfo{title}{{Robust Machine Learning Applied to Astronomical Data
  Sets. I. Star-Galaxy Classification of the Sloan Digital Sky Survey DR3 Using
  Decision Trees}}.
\newblock \bibinfo{journal}{The Astrophysical Journal} \bibinfo{volume}{650},
  \bibinfo{pages}{497--509}.
\newblock \DOIprefix\doi{10.1086/507440},
  \href{http://arxiv.org/abs/astro-ph/0606541}{\tt arXiv:astro-ph/0606541}.
\bibitem[{{Bertin} and {Arnouts}(1996)}]{bert96}
\bibinfo{author}{{Bertin}, E.}, \bibinfo{author}{{Arnouts}, S.},
  \bibinfo{year}{1996}.
\newblock \bibinfo{title}{{SExtractor: Software for source extraction.}}
\newblock \bibinfo{journal}{Astronomy \& Astrophysics} \bibinfo{volume}{117},
  \bibinfo{pages}{393--404}.
\bibitem[{{Brun} and {Rademakers}(1997)}]{root96}
\bibinfo{author}{{Brun}, R.}, \bibinfo{author}{{Rademakers}, F.},
  \bibinfo{year}{1997}.
\newblock \bibinfo{title}{{ROOT: An object oriented data analysis framework}}.
\newblock \bibinfo{journal}{Nuclear Instruments and Methods in Physics Research
  A} \bibinfo{volume}{389}, \bibinfo{pages}{81--86}.
\newblock \DOIprefix\doi{10.1016/S0168-9002(97)00048-X}.
\bibitem[{{Carrasco Kind} and {Brunner}(2013)}]{carr13}
\bibinfo{author}{{Carrasco Kind}, M.}, \bibinfo{author}{{Brunner}, R.J.},
  \bibinfo{year}{2013}.
\newblock \bibinfo{title}{{TPZ: photometric redshift PDFs and ancillary
  information by using prediction trees and random forests}}.
\newblock \bibinfo{journal}{Monthly Notices of the Royal Astronomical Society}
  \bibinfo{volume}{432}, \bibinfo{pages}{1483--1501}.
\newblock \DOIprefix\doi{10.1093/mnras/stt574},
  \href{http://arxiv.org/abs/1303.7269}{\tt arXiv:1303.7269}.
\bibitem[{{Carrasco Kind} and {Brunner}(2014)}]{carr14}
\bibinfo{author}{{Carrasco Kind}, M.}, \bibinfo{author}{{Brunner}, R.J.},
  \bibinfo{year}{2014}.
\newblock \bibinfo{title}{{Exhausting the information: novel Bayesian
  combination of photometric redshift PDFs}}.
\newblock \bibinfo{journal}{Monthly Notices of the Royal Astronomical Society}
  \bibinfo{volume}{442}, \bibinfo{pages}{3380--3399}.
\newblock \DOIprefix\doi{10.1093/mnras/stu1098},
  \href{http://arxiv.org/abs/1403.0044}{\tt arXiv:1403.0044}.
\bibitem[{CMS-Collaboration(2012)}]{cms12}
\bibinfo{author}{CMS-Collaboration}, \bibinfo{year}{2012}.
\newblock \bibinfo{title}{Observation of a new boson at a mass of 125 gev with
  the \{CMS\} experiment at the \{LHC\}}.
\newblock \bibinfo{journal}{Physics Letters B} \bibinfo{volume}{716},
  \bibinfo{pages}{30 -- 61}.
\newblock \URLprefix
  \url{http://www.sciencedirect.com/science/article/pii/S0370269312008581},
  \DOIprefix\doi{http://dx.doi.org/10.1016/j.physletb.2012.08.021}.
\bibitem[{{Coupon} et~al.(2012){Coupon}, {Kilbinger}, {McCracken}, {Ilbert},
  {Arnouts}, {Mellier}, {Abbas}, {de la Torre}, {Goranova}, {Hudelot}, {Kneib}
  and {Le F{\`e}vre}}]{coup12}
\bibinfo{author}{{Coupon}, J.}, \bibinfo{author}{{Kilbinger}, M.},
  \bibinfo{author}{{McCracken}, H.J.}, \bibinfo{author}{{Ilbert}, O.},
  \bibinfo{author}{{Arnouts}, S.}, \bibinfo{author}{{Mellier}, Y.},
  \bibinfo{author}{{Abbas}, U.}, \bibinfo{author}{{de la Torre}, S.},
  \bibinfo{author}{{Goranova}, Y.}, \bibinfo{author}{{Hudelot}, P.},
  \bibinfo{author}{{Kneib}, J.P.}, \bibinfo{author}{{Le F{\`e}vre}, O.},
  \bibinfo{year}{2012}.
\newblock \bibinfo{title}{{Galaxy clustering in the CFHTLS-Wide: the changing
  relationship between galaxies and haloes since $z = 1.2$}}.
\newblock \bibinfo{journal}{Astronomy \& Astrophysics} \bibinfo{volume}{542},
  \bibinfo{pages}{A5}.
\newblock \DOIprefix\doi{10.1051/0004-6361/201117625},
  \href{http://arxiv.org/abs/1107.0616}{\tt arXiv:1107.0616}.
\bibitem[{{Desai} et~al.(2012){Desai}, {Armstrong}, {Mohr}, {Semler}, {Liu},
  {Bertin}, {Allam}, {Barkhouse}, {Bazin}, {Buckley-Geer}, {Cooper}, {Hansen},
  {High}, {Lin}, {Lin}, {Ngeow}, {Rest}, {Song}, {Tucker} and
  {Zenteno}}]{desa12}
\bibinfo{author}{{Desai}, S.}, \bibinfo{author}{{Armstrong}, R.},
  \bibinfo{author}{{Mohr}, J.J.}, \bibinfo{author}{{Semler}, D.R.},
  \bibinfo{author}{{Liu}, J.}, \bibinfo{author}{{Bertin}, E.},
  \bibinfo{author}{{Allam}, S.S.}, \bibinfo{author}{{Barkhouse}, W.A.},
  \bibinfo{author}{{Bazin}, G.}, \bibinfo{author}{{Buckley-Geer}, E.J.},
  \bibinfo{author}{{Cooper}, M.C.}, \bibinfo{author}{{Hansen}, S.M.},
  \bibinfo{author}{{High}, F.W.}, \bibinfo{author}{{Lin}, H.},
  \bibinfo{author}{{Lin}, Y.T.}, \bibinfo{author}{{Ngeow}, C.C.},
  \bibinfo{author}{{Rest}, A.}, \bibinfo{author}{{Song}, J.},
  \bibinfo{author}{{Tucker}, D.}, \bibinfo{author}{{Zenteno}, A.},
  \bibinfo{year}{2012}.
\newblock \bibinfo{title}{{The Blanco Cosmology Survey: Data Acquisition,
  Processing, Calibration, Quality Diagnostics, and Data Release}}.
\newblock \bibinfo{journal}{The Astrophysical Journal} \bibinfo{volume}{757},
  \bibinfo{pages}{83}.
\newblock \DOIprefix\doi{10.1088/0004-637X/757/1/83},
  \href{http://arxiv.org/abs/1204.1210}{\tt arXiv:1204.1210}.
\bibitem[{{Etayo-Sotos} and {Sevilla-Noarbe}(2013)}]{etay12}
\bibinfo{author}{{Etayo-Sotos}, P.}, \bibinfo{author}{{Sevilla-Noarbe}, I.},
  \bibinfo{year}{2013}.
\newblock \bibinfo{title}{{Using boosted decision trees for star-galaxy
  separation}}, in: \bibinfo{editor}{{Guirado}, J.C.}, \bibinfo{editor}{{Lara},
  L.M.}, \bibinfo{editor}{{Quilis}, V.}, \bibinfo{editor}{{Gorgas}, J.} (Eds.),
  \bibinfo{booktitle}{Highlights of Spanish Astrophysics VII}, pp.
  \bibinfo{pages}{944--944}.
\bibitem[{{Fadely} et~al.(2012){Fadely}, {Hogg} and {Willman}}]{fade12}
\bibinfo{author}{{Fadely}, R.}, \bibinfo{author}{{Hogg}, D.W.},
  \bibinfo{author}{{Willman}, B.}, \bibinfo{year}{2012}.
\newblock \bibinfo{title}{{Star-Galaxy Classification in Multi-band Optical
  Imaging}}.
\newblock \bibinfo{journal}{The Astrophysical Journal} \bibinfo{volume}{760},
  \bibinfo{pages}{15}.
\newblock \DOIprefix\doi{10.1088/0004-637X/760/1/15},
  \href{http://arxiv.org/abs/1206.4306}{\tt arXiv:1206.4306}.
\bibitem[{Fermi-LAT-Collaboration(2012)}]{acke12}
\bibinfo{author}{Fermi-LAT-Collaboration}, \bibinfo{year}{2012}.
\newblock \bibinfo{title}{{A Statistical Approach to Recognizing Source Classes
  for Unassociated Sources in the First Fermi-LAT Catalog}}.
\newblock \bibinfo{journal}{The Astrophysical Journal} \bibinfo{volume}{753},
  \bibinfo{pages}{83}.
\newblock \DOIprefix\doi{10.1088/0004-637X/753/1/83},
  \href{http://arxiv.org/abs/1108.1202}{\tt arXiv:1108.1202}.
\bibitem[{Freund and Schapire(1997)}]{freu97}
\bibinfo{author}{Freund, Y.}, \bibinfo{author}{Schapire, R.E.},
  \bibinfo{year}{1997}.
\newblock \bibinfo{title}{A decision-theoretic generalization of on-line
  learning and an application to boosting}.
\newblock \bibinfo{journal}{Journal of Computer and System Sciences}
  \bibinfo{volume}{55}, \bibinfo{pages}{119 -- 139}.
\newblock \URLprefix
  \url{http://www.sciencedirect.com/science/article/pii/S002200009791504X},
  \DOIprefix\doi{http://dx.doi.org/10.1006/jcss.1997.1504}.
\bibitem[{{Gerdes} et~al.(2010){Gerdes}, {Sypniewski}, {McKay}, {Hao}, {Weis},
  {Wechsler} and {Busha}}]{gerd10}
\bibinfo{author}{{Gerdes}, D.W.}, \bibinfo{author}{{Sypniewski}, A.J.},
  \bibinfo{author}{{McKay}, T.A.}, \bibinfo{author}{{Hao}, J.},
  \bibinfo{author}{{Weis}, M.R.}, \bibinfo{author}{{Wechsler}, R.H.},
  \bibinfo{author}{{Busha}, M.T.}, \bibinfo{year}{2010}.
\newblock \bibinfo{title}{{ArborZ: Photometric Redshifts Using Boosted Decision
  Trees}}.
\newblock \bibinfo{journal}{The Astrophysical Journal} \bibinfo{volume}{715},
  \bibinfo{pages}{823--832}.
\newblock \DOIprefix\doi{10.1088/0004-637X/715/2/823},
  \href{http://arxiv.org/abs/0908.4085}{\tt arXiv:0908.4085}.
\bibitem[{{Giannantonio} et~al.(2014){Giannantonio}, {Ross}, {Percival},
  {Crittenden}, {Bacher}, {Kilbinger}, {Nichol} and {Weller}}]{gian14}
\bibinfo{author}{{Giannantonio}, T.}, \bibinfo{author}{{Ross}, A.J.},
  \bibinfo{author}{{Percival}, W.J.}, \bibinfo{author}{{Crittenden}, R.},
  \bibinfo{author}{{Bacher}, D.}, \bibinfo{author}{{Kilbinger}, M.},
  \bibinfo{author}{{Nichol}, R.}, \bibinfo{author}{{Weller}, J.},
  \bibinfo{year}{2014}.
\newblock \bibinfo{title}{{Improved primordial non-Gaussianity constraints from
  measurements of galaxy clustering and the integrated Sachs-Wolfe effect}}.
\newblock \bibinfo{journal}{Physical Review D} \bibinfo{volume}{89},
  \bibinfo{pages}{023511}.
\newblock \DOIprefix\doi{10.1103/PhysRevD.89.023511},
  \href{http://arxiv.org/abs/1303.1349}{\tt arXiv:1303.1349}.
\bibitem[{{Heydon-Dumbleton} et~al.(1989){Heydon-Dumbleton}, {Collins} and
  {MacGillivray}}]{heyd89}
\bibinfo{author}{{Heydon-Dumbleton}, N.H.}, \bibinfo{author}{{Collins}, C.A.},
  \bibinfo{author}{{MacGillivray}, H.T.}, \bibinfo{year}{1989}.
\newblock \bibinfo{title}{{The Edinburgh/Durham Southern Galaxy Catalogue. II -
  Image classification and galaxy number counts}}.
\newblock \bibinfo{journal}{Monthly Notices of the Royal Astronomical Society}
  \bibinfo{volume}{238}, \bibinfo{pages}{379--406}.
\bibitem[{Hildebrandt et~al.(2012)Hildebrandt, Erben, Kuijken, van Waerbeke,
  Heymans et~al.}]{hild12}
\bibinfo{author}{Hildebrandt, H.}, \bibinfo{author}{Erben, T.},
  \bibinfo{author}{Kuijken, K.}, \bibinfo{author}{van Waerbeke, L.},
  \bibinfo{author}{Heymans, C.}, et~al., \bibinfo{year}{2012}.
\newblock \bibinfo{title}{{CFHTLenS: Improving the quality of photometric
  redshifts with precision photometry}}.
\newblock \bibinfo{journal}{Mon.Not.Roy.Astron.Soc.} \bibinfo{volume}{421},
  \bibinfo{pages}{2355}.
\newblock \DOIprefix\doi{10.1111/j.1365-2966.2012.20468.x},
  \href{http://arxiv.org/abs/1111.4434}{\tt arXiv:1111.4434}.
\bibitem[{Hoecker et~al.(2007)Hoecker, Speckmayer, Stelzer, Therhaag, von
  Toerne and Voss}]{tmva07}
\bibinfo{author}{Hoecker, A.}, \bibinfo{author}{Speckmayer, P.},
  \bibinfo{author}{Stelzer, J.}, \bibinfo{author}{Therhaag, J.},
  \bibinfo{author}{von Toerne, E.}, \bibinfo{author}{Voss, H.},
  \bibinfo{year}{2007}.
\newblock \bibinfo{title}{{TMVA: Toolkit for Multivariate Data Analysis}}.
\newblock \bibinfo{journal}{PoS} \bibinfo{volume}{ACAT}, \bibinfo{pages}{040}.
\newblock \href{http://arxiv.org/abs/physics/0703039}{\tt
  arXiv:physics/0703039}.
\bibitem[{{MacGillivray} et~al.(1976){MacGillivray}, {Martin}, {Pratt},
  {Reddish}, {Seddon}, {Alexander}, {Walker} and {Williams}}]{macg76}
\bibinfo{author}{{MacGillivray}, H.T.}, \bibinfo{author}{{Martin}, R.},
  \bibinfo{author}{{Pratt}, N.M.}, \bibinfo{author}{{Reddish}, V.C.},
  \bibinfo{author}{{Seddon}, H.}, \bibinfo{author}{{Alexander}, L.W.G.},
  \bibinfo{author}{{Walker}, G.S.}, \bibinfo{author}{{Williams}, P.R.},
  \bibinfo{year}{1976}.
\newblock \bibinfo{title}{{A method for the automatic separation of the images
  of galaxies and stars from measurements made with the COSMOS machine}}.
\newblock \bibinfo{journal}{Monthly Notices of the Royal Astronomical Society}
  \bibinfo{volume}{176}, \bibinfo{pages}{265--274}.
\bibitem[{{Maddox} et~al.(1990){Maddox}, {Efstathiou}, {Sutherland} and
  {Loveday}}]{madd90}
\bibinfo{author}{{Maddox}, S.J.}, \bibinfo{author}{{Efstathiou}, G.},
  \bibinfo{author}{{Sutherland}, W.J.}, \bibinfo{author}{{Loveday}, J.},
  \bibinfo{year}{1990}.
\newblock \bibinfo{title}{{The APM galaxy survey. I - APM measurements and
  star-galaxy separation}}.
\newblock \bibinfo{journal}{Monthly Notices of the Royal Astronomical Society}
  \bibinfo{volume}{243}, \bibinfo{pages}{692--712}.
\bibitem[{{Ma{\l}ek} et~al.(2013){Ma{\l}ek}, {Solarz}, {Pollo}, {Fritz},
  {Garilli}, {Scodeggio}, {Iovino}, {Granett}, {Abbas}, {Adami}, {Arnouts},
  {Bel}, {Bolzonella}, {Bottini}, {Branchini}, {Cappi}, {Coupon}, {Cucciati},
  {Davidzon}, {De Lucia}, {de la Torre}, {Franzetti}, {Fumana}, {Guzzo},
  {Ilbert}, {Krywult}, {Le Brun}, {Le Fevre}, {Maccagni}, {Marulli},
  {McCracken}, {Paioro}, {Polletta}, {Schlagenhaufer}, {Tasca}, {Tojeiro},
  {Vergani}, {Zanichelli}, {Burden}, {Di Porto}, {Marchetti}, {Marinoni},
  {Mellier}, {Moscardini}, {Nichol}, {Peacock}, {Percival}, {Phleps}, {Wolk}
  and {Zamorani}}]{male13}
\bibinfo{author}{{Ma{\l}ek}, K.}, \bibinfo{author}{{Solarz}, A.},
  \bibinfo{author}{{Pollo}, A.}, \bibinfo{author}{{Fritz}, A.},
  \bibinfo{author}{{Garilli}, B.}, \bibinfo{author}{{Scodeggio}, M.},
  \bibinfo{author}{{Iovino}, A.}, \bibinfo{author}{{Granett}, B.R.},
  \bibinfo{author}{{Abbas}, U.}, \bibinfo{author}{{Adami}, C.},
  \bibinfo{author}{{Arnouts}, S.}, \bibinfo{author}{{Bel}, J.},
  \bibinfo{author}{{Bolzonella}, M.}, \bibinfo{author}{{Bottini}, D.},
  \bibinfo{author}{{Branchini}, E.}, \bibinfo{author}{{Cappi}, A.},
  \bibinfo{author}{{Coupon}, J.}, \bibinfo{author}{{Cucciati}, O.},
  \bibinfo{author}{{Davidzon}, I.}, \bibinfo{author}{{De Lucia}, G.},
  \bibinfo{author}{{de la Torre}, S.}, \bibinfo{author}{{Franzetti}, P.},
  \bibinfo{author}{{Fumana}, M.}, \bibinfo{author}{{Guzzo}, L.},
  \bibinfo{author}{{Ilbert}, O.}, \bibinfo{author}{{Krywult}, J.},
  \bibinfo{author}{{Le Brun}, V.}, \bibinfo{author}{{Le Fevre}, O.},
  \bibinfo{author}{{Maccagni}, D.}, \bibinfo{author}{{Marulli}, F.},
  \bibinfo{author}{{McCracken}, H.J.}, \bibinfo{author}{{Paioro}, L.},
  \bibinfo{author}{{Polletta}, M.}, \bibinfo{author}{{Schlagenhaufer}, H.},
  \bibinfo{author}{{Tasca}, L.A.M.}, \bibinfo{author}{{Tojeiro}, R.},
  \bibinfo{author}{{Vergani}, D.}, \bibinfo{author}{{Zanichelli}, A.},
  \bibinfo{author}{{Burden}, A.}, \bibinfo{author}{{Di Porto}, C.},
  \bibinfo{author}{{Marchetti}, A.}, \bibinfo{author}{{Marinoni}, C.},
  \bibinfo{author}{{Mellier}, Y.}, \bibinfo{author}{{Moscardini}, L.},
  \bibinfo{author}{{Nichol}, R.C.}, \bibinfo{author}{{Peacock}, J.A.},
  \bibinfo{author}{{Percival}, W.J.}, \bibinfo{author}{{Phleps}, S.},
  \bibinfo{author}{{Wolk}, M.}, \bibinfo{author}{{Zamorani}, G.},
  \bibinfo{year}{2013}.
\newblock \bibinfo{title}{{The VIMOS Public Extragalactic Redshift Survey
  (VIPERS). A support vector machine classification of galaxies, stars, and
  AGNs}}.
\newblock \bibinfo{journal}{Astronomy \& Astrophysics} \bibinfo{volume}{557},
  \bibinfo{pages}{A16}.
\newblock \DOIprefix\doi{10.1051/0004-6361/201321447},
  \href{http://arxiv.org/abs/1303.2621}{\tt arXiv:1303.2621}.
\bibitem[{{Miller} and {Coe}(1996)}]{mill96}
\bibinfo{author}{{Miller}, A.S.}, \bibinfo{author}{{Coe}, M.J.},
  \bibinfo{year}{1996}.
\newblock \bibinfo{title}{{Star/galaxy classification using Kohonen
  self-organizing maps}}.
\newblock \bibinfo{journal}{Monthly Notices of the Royal Astronomical Society}
  \bibinfo{volume}{279}, \bibinfo{pages}{293--300}.
\bibitem[{{Odewahn} et~al.(1992){Odewahn}, {Stockwell}, {Pennington},
  {Humphreys} and {Zumach}}]{odew92}
\bibinfo{author}{{Odewahn}, S.C.}, \bibinfo{author}{{Stockwell}, E.B.},
  \bibinfo{author}{{Pennington}, R.L.}, \bibinfo{author}{{Humphreys}, R.M.},
  \bibinfo{author}{{Zumach}, W.A.}, \bibinfo{year}{1992}.
\newblock \bibinfo{title}{{Automated star/galaxy discrimination with neural
  networks}}.
\newblock \bibinfo{journal}{Astronomical Journal} \bibinfo{volume}{103},
  \bibinfo{pages}{318--331}.
\newblock \DOIprefix\doi{10.1086/116063}.
\bibitem[{Roe et~al.(2005)Roe, Yang, Zhu, Liu, Stancu and McGregor}]{roe05}
\bibinfo{author}{Roe, B.P.}, \bibinfo{author}{Yang, H.J.},
  \bibinfo{author}{Zhu, J.}, \bibinfo{author}{Liu, Y.},
  \bibinfo{author}{Stancu, I.}, \bibinfo{author}{McGregor, G.},
  \bibinfo{year}{2005}.
\newblock \bibinfo{title}{Boosted decision trees as an alternative to
  artificial neural networks for particle identification}.
\newblock \bibinfo{journal}{Nuclear Instruments and Methods in Physics Research
  Section A: Accelerators, Spectrometers, Detectors and Associated Equipment}
  \bibinfo{volume}{543}, \bibinfo{pages}{577 -- 584}.
\newblock \URLprefix
  \url{http://www.sciencedirect.com/science/article/pii/S0168900205000355},
  \DOIprefix\doi{http://dx.doi.org/10.1016/j.nima.2004.12.018}.
\bibitem[{{Ross} et~al.(2011){Ross}, {Ho}, {Cuesta}, {Tojeiro}, {Percival},
  {Wake}, {Masters}, {Nichol}, {Myers}, {de Simoni}, {Seo},
  {Hern{\'a}ndez-Monteagudo}, {Crittenden}, {Blanton}, {Brinkmann}, {da Costa},
  {Guo}, {Kazin}, {Maia}, {Maraston}, {Padmanabhan}, {Prada}, {Ramos},
  {Sanchez}, {Schlafly}, {Schlegel}, {Schneider}, {Skibba}, {Thomas}, {Weaver},
  {White} and {Zehavi}}]{ross11}
\bibinfo{author}{{Ross}, A.J.}, \bibinfo{author}{{Ho}, S.},
  \bibinfo{author}{{Cuesta}, A.J.}, \bibinfo{author}{{Tojeiro}, R.},
  \bibinfo{author}{{Percival}, W.J.}, \bibinfo{author}{{Wake}, D.},
  \bibinfo{author}{{Masters}, K.L.}, \bibinfo{author}{{Nichol}, R.C.},
  \bibinfo{author}{{Myers}, A.D.}, \bibinfo{author}{{de Simoni}, F.},
  \bibinfo{author}{{Seo}, H.J.}, \bibinfo{author}{{Hern{\'a}ndez-Monteagudo},
  C.}, \bibinfo{author}{{Crittenden}, R.}, \bibinfo{author}{{Blanton}, M.},
  \bibinfo{author}{{Brinkmann}, J.}, \bibinfo{author}{{da Costa}, L.A.N.},
  \bibinfo{author}{{Guo}, H.}, \bibinfo{author}{{Kazin}, E.},
  \bibinfo{author}{{Maia}, M.A.G.}, \bibinfo{author}{{Maraston}, C.},
  \bibinfo{author}{{Padmanabhan}, N.}, \bibinfo{author}{{Prada}, F.},
  \bibinfo{author}{{Ramos}, B.}, \bibinfo{author}{{Sanchez}, A.},
  \bibinfo{author}{{Schlafly}, E.F.}, \bibinfo{author}{{Schlegel}, D.J.},
  \bibinfo{author}{{Schneider}, D.P.}, \bibinfo{author}{{Skibba}, R.},
  \bibinfo{author}{{Thomas}, D.}, \bibinfo{author}{{Weaver}, B.A.},
  \bibinfo{author}{{White}, M.}, \bibinfo{author}{{Zehavi}, I.},
  \bibinfo{year}{2011}.
\newblock \bibinfo{title}{{Ameliorating systematic uncertainties in the angular
  clustering of galaxies: a study using the SDSS-III}}.
\newblock \bibinfo{journal}{Monthly Notices of the Royal Astronomical Society}
  \bibinfo{volume}{417}, \bibinfo{pages}{1350--1373}.
\newblock \DOIprefix\doi{10.1111/j.1365-2966.2011.19351.x},
  \href{http://arxiv.org/abs/1105.2320}{\tt arXiv:1105.2320}.
\bibitem[{{Sebok}(1979)}]{sebo79}
\bibinfo{author}{{Sebok}, W.L.}, \bibinfo{year}{1979}.
\newblock \bibinfo{title}{{Optimal classification of images into stars or
  galaxies - A Bayesian approach}}.
\newblock \bibinfo{journal}{Astronomical Journal} \bibinfo{volume}{84},
  \bibinfo{pages}{1526--1536}.
\newblock \DOIprefix\doi{10.1086/112570}.
\bibitem[{{Soumagnac} et~al.(2015){Soumagnac}, {Abdalla}, {Lahav}, {Kirk},
  {Sevilla}, {Bertin}, {Rowe}, {Annis}, {Busha}, {Da Costa}, {Frieman},
  {Gaztanaga}, {Jarvis}, {Lin}, {Percival}, {Santiago}, {Sabiu}, {Wechsler},
  {Wolz} and {Yanny}}]{soum13}
\bibinfo{author}{{Soumagnac}, M.T.}, \bibinfo{author}{{Abdalla}, F.B.},
  \bibinfo{author}{{Lahav}, O.}, \bibinfo{author}{{Kirk}, D.},
  \bibinfo{author}{{Sevilla}, I.}, \bibinfo{author}{{Bertin}, E.},
  \bibinfo{author}{{Rowe}, B.T.P.}, \bibinfo{author}{{Annis}, J.},
  \bibinfo{author}{{Busha}, M.T.}, \bibinfo{author}{{Da Costa}, L.N.},
  \bibinfo{author}{{Frieman}, J.A.}, \bibinfo{author}{{Gaztanaga}, E.},
  \bibinfo{author}{{Jarvis}, M.}, \bibinfo{author}{{Lin}, H.},
  \bibinfo{author}{{Percival}, W.J.}, \bibinfo{author}{{Santiago}, B.X.},
  \bibinfo{author}{{Sabiu}, C.G.}, \bibinfo{author}{{Wechsler}, R.H.},
  \bibinfo{author}{{Wolz}, L.}, \bibinfo{author}{{Yanny}, B.},
  \bibinfo{year}{2015}.
\newblock \bibinfo{title}{{Star/galaxy separation at faint magnitudes:
  application to a simulated Dark Energy Survey}}.
\newblock \bibinfo{journal}{Monthly Notices of the Royal Astronomical Society}
  \bibinfo{volume}{450}, \bibinfo{pages}{666--680}.
\newblock \DOIprefix\doi{10.1093/mnras/stu1410},
  \href{http://arxiv.org/abs/1306.5236}{\tt arXiv:1306.5236}.
\bibitem[{{Suchkov} et~al.(2005){Suchkov}, {Hanisch} and {Margon}}]{such05}
\bibinfo{author}{{Suchkov}, A.A.}, \bibinfo{author}{{Hanisch}, R.J.},
  \bibinfo{author}{{Margon}, B.}, \bibinfo{year}{2005}.
\newblock \bibinfo{title}{{A Census of Object Types and Redshift Estimates in
  the SDSS Photometric Catalog from a Trained Decision Tree Classifier}}.
\newblock \bibinfo{journal}{Astronomical Journal} \bibinfo{volume}{130},
  \bibinfo{pages}{2439--2452}.
\newblock \DOIprefix\doi{10.1086/497363},
  \href{http://arxiv.org/abs/astro-ph/0508501}{\tt arXiv:astro-ph/0508501}.
\bibitem[{{Thomas} et~al.(2011){Thomas}, {Abdalla} and {Lahav}}]{thom11}
\bibinfo{author}{{Thomas}, S.A.}, \bibinfo{author}{{Abdalla}, F.B.},
  \bibinfo{author}{{Lahav}, O.}, \bibinfo{year}{2011}.
\newblock \bibinfo{title}{{Excess Clustering on Large Scales in the MegaZ DR7
  Photometric Redshift Survey}}.
\newblock \bibinfo{journal}{Physical Review Letters} \bibinfo{volume}{106},
  \bibinfo{pages}{241301}.
\newblock \DOIprefix\doi{10.1103/PhysRevLett.106.241301},
  \href{http://arxiv.org/abs/1012.2272}{\tt arXiv:1012.2272}.
\bibitem[{{Vasconcellos} et~al.(2011){Vasconcellos}, {de Carvalho}, {Gal},
  {LaBarbera}, {Capelato}, {Frago Campos Velho}, {Trevisan} and
  {Ruiz}}]{vasc11}
\bibinfo{author}{{Vasconcellos}, E.C.}, \bibinfo{author}{{de Carvalho}, R.R.},
  \bibinfo{author}{{Gal}, R.R.}, \bibinfo{author}{{LaBarbera}, F.L.},
  \bibinfo{author}{{Capelato}, H.V.}, \bibinfo{author}{{Frago Campos Velho},
  H.}, \bibinfo{author}{{Trevisan}, M.}, \bibinfo{author}{{Ruiz}, R.S.R.},
  \bibinfo{year}{2011}.
\newblock \bibinfo{title}{{Decision Tree Classifiers for Star/Galaxy
  Separation}}.
\newblock \bibinfo{journal}{Astronomical Journal} \bibinfo{volume}{141},
  \bibinfo{pages}{189}.
\newblock \DOIprefix\doi{10.1088/0004-6256/141/6/189},
  \href{http://arxiv.org/abs/1011.1951}{\tt arXiv:1011.1951}.
\bibitem[{{Weir} et~al.(1995){Weir}, {Fayyad} and {Djorgovski}}]{weir95}
\bibinfo{author}{{Weir}, N.}, \bibinfo{author}{{Fayyad}, U.M.},
  \bibinfo{author}{{Djorgovski}, S.}, \bibinfo{year}{1995}.
\newblock \bibinfo{title}{{Automated Star/Galaxy Classification for Digitized
  Poss-II}}.
\newblock \bibinfo{journal}{Astronomical Journal} \bibinfo{volume}{109},
  \bibinfo{pages}{2401}.
\newblock \DOIprefix\doi{10.1086/117459}.
\bibitem[{Yang et~al.(2005)Yang, Roe and Zhu}]{yang05}
\bibinfo{author}{Yang, H.J.}, \bibinfo{author}{Roe, B.P.},
  \bibinfo{author}{Zhu, J.}, \bibinfo{year}{2005}.
\newblock \bibinfo{title}{Studies of boosted decision trees for miniboone
  particle identification}.
\newblock \bibinfo{journal}{Nuclear Instruments and Methods in Physics Research
  Section A: Accelerators, Spectrometers, Detectors and Associated Equipment}
  \bibinfo{volume}{555}, \bibinfo{pages}{370 -- 385}.
\newblock \URLprefix
  \url{http://www.sciencedirect.com/science/article/pii/S0168900205018322},
  \DOIprefix\doi{http://dx.doi.org/10.1016/j.nima.2005.09.022}.

\end{thebibliography}

\clearpage

\end{document}